\documentclass[twocolumn,amsmath,amssymb,aps,pra]{revtex4-2}
\usepackage{graphicx}% Include figure files
\usepackage{dcolumn}% Align table columns on decimal point
\usepackage{bm}% bold math
\usepackage{color}
\begin{document}

\title{Chaos in coupled Kerr-nonlinear parametric oscillators}

\author{Hayato Goto and Taro Kanao}
\affiliation{
Frontier Research Laboratory, 
Corporate Research \& Development Center, 
Toshiba Corporation, 
1, Komukai Toshiba-cho, Saiwai-ku, Kawasaki-shi, 212-8582, Japan}

\date{\today}

\begin{abstract}

A Kerr-nonlinear parametric oscillator (KPO) can generate a quantum superposition of two oscillating states,
known as a Schr\"{o}dinger cat state, via quantum adiabatic evolution, 
and can be used as a qubit for gate-based quantum computing and quantum annealing.
In this work, we investigate complex dynamics, i.e., chaos, in two coupled nondissipative KPOs at a few-photon level.
After showing that a classical model for this system is nonintegrable and consequently exhibits chaotic behavior,
we provide quantum counterparts for the classical results, which are 
quantum versions of the Poincar\'{e} surface of section and its lower-dimensional version 
defined with time integrals of the Wigner and Husimi functions, and also 
the initial and long-term behavior of out-of-time-ordered correlators.
We conclude that some of them can be regarded as quantum signatures of chaos,
together with energy-level spacing statistics (conventional signature).
Thus, the system of coupled KPOs is expected to offer not only an alternative approach to quantum computing, 
but also a promising platform for the study on quantum chaos.

\end{abstract}

\maketitle

\section{Introduction}

Computational basis states for quantum computing are usually taken from quantized energy levels.
An alternative approach to it is to use oscillator states consisting of multiple quanta (photons or phonons).
In this approach, multiple number states are used to represent each basis state, like logical qubits 
for quantum error correction~\cite{Nielsen}.
Thus, the oscillator approach to quantum computing will offer qubits robust to noises or 
hardware-efficient quantum error-correcting codes, which are known as 
bosonic codes~\cite{Gottesman2001a,Cochrane1999a,Michael2016a,Ofek2016a,Chou2018a,Hu2019a,Fluhmann2019a,Campagne2020a}.

One of such oscillator approaches is to use parametric oscillators. 
Their two stable oscillating states can be used for qubits. 
There are two types of parametric-oscillator approach:
dissipative and Kerr.
In the dissipative-type approach,
parametrically two-photon driven oscillators with large two-photon loss (larger than single-photon loss)
are used for qubits~\cite{Mirrahimi2014a,Albert2016a}.
Such a two-photon dissipative parametric oscillator becomes a Schr\"{o}dinger cat state 
(superposition of two oscillating states) as a steady state~\cite{Mirrahimi2014a,Albert2016a,Milburn},
which has been demonstrated experimentally using superconducting circuits~\cite{Leghtas2015a,Touzard2018a,Lescanne2020a}.
Since this type of qubit is insensitive to bit-flip errors~\cite{Lescanne2020a}, 
this approach is expected to be useful for fault-tolerant quantum computation~\cite{Guillaud2019a,Chamberland2020a}.

In the Kerr-type approach, 
parametrically two-photon driven oscillators with large Kerr nonlinearity~\cite{Milburn1991a,Wielinga1993a}, 
which we call Kerr-nonlinear parametric oscillators (KPOs)~\cite{Goto2016a,Goto2019a}, 
are used for qubits.
Low-loss KPOs have recently been realized experimentally using 
superconducting circuits~\cite{Wang2019a,Grimm2020a,Yamaji2020a}.
An ideal KPO is lossless (nondissipative), and it can generate a Schr\"{o}dinger cat state from the vacuum state 
via quantum adiabatic evolution (quantum bifurcation)~\cite{Cochrane1999a,Goto2016a}.
Moreover, a network of KPOs can solve a combinatorial optimization problem 
(ground-state search in the Ising model) 
by adiabatic quantum computation~\cite{Farhi2000a,Farhi2001a,Albash2018a} or 
quantum annealing~\cite{Kadowaki1998a,Das2008a}, 
the final state of which is a highly entangled state, a superposition of many-mode coherent states 
corresponding to two optimal solutions~\cite{Goto2016a}.
Quantum annealing using KPOs has been developed 
in this five years~\cite{Nigg2017a,Puri2017a,Zhao2018a,Onodera2020a,Goto2020a,Kanao2021a}.
The KPOs can also be used for qubits in gate-based quantum computing~\cite{Goto2019a,Goto2016b,Puri2017b,Puri2020a,Kanao2021b,Xu2021a}.
Since the KPO qubit, also known as a Kerr-cat qubit, is robust against bit-flip errors, like the above dissipative-type qubit, 
fault-tolerant quantum computation using KPOs has been developed~\cite{Puri2019a,Darmawan2021a}.
The KPO has also offered physically interesting topics, such as 
nonclassical traveling-state generation~\cite{Goto2019b,Strandberg2021a}, 
quantum heating leading to Boltzmann sampling~\cite{Goto2018a},
steady-state entanglement generation~\cite{Manaev2018a,Kewming2020a}, and 
phase transition~\cite{Savona2017a,Rota2019a}.

In this work, we investigate nonlinear dynamics of coupled KPOs from the viewpoint of chaos~\cite{Strogatz,Wimberger}.
The KPO, which is nondissipative in an ideal case, is more desirable for the study on chaos 
than the two-photon dissipative parametric oscillator
(and also optical parametric oscillators~\cite{Goto2019a,Wang2013a}, another dissipative type), 
because dissipation inevitably introduces noises (so-called quantum noises)~\cite{Milburn,Gardiner,Breuer}, 
and such stochastic noises are undesirable for the study on chaos~\cite{Strogatz}.
Nonlinear dynamics of a KPO have been studied~\cite{Milburn1991a,Hovsepyan2016a}. 
To our knowledge, however, the coupled-KPO case has not been explored so far, 
though it has been suggested that the studies on chaos in a KPO network would be interesting~\cite{Goto2016a}.
(Chaos in a simplified classical model for the KPO network has also been suggested in the proposal of 
a quantum-inspired algorithm called simulated bifurcation~\cite{Goto2019c}, but it has not been investigated in detail.)
Here we treat a system of two nondissipative KPOs with time-independent parameters as the simplest example 
sufficient for the study on chaos.
(Similar studies on more KPOs are an interesting next step.)

This paper is organized as follows. 
In Sec.~\ref{sec-model}, we introduce the quantum and classical models for the system. 
In Sec.~\ref{sec-classical}, we show our results for the classical model, 
where the nonintegrability of this model is shown 
by the Poincar\'{e} surface of section (SOS)~\cite{Strogatz,Wimberger} and its lower-dimensional version, and 
the sensitivity to initial conditions is also shown. 
These indicate chaos in the classical model.
In Sec.~\ref{sec-quantum}, we provide our results for the quantum model, 
where the SOS and its lower-dimensional version are extended to quantum cases using the Wigner and Husimi functions~\cite{Milburn,Wimberger,Leonhardt}, 
and the initial-condition sensitivity is also examined using 
out-of-time-ordered correlators (OTOCs)~\cite{Hashimoto2017a,Fortes2019a,Akutagawa2020a,Hashimoto2020a,Bhattacharyya2021a}.
Energy-level spacing statistics, 
which is a conventional quantum signature of chaos~\cite{Wimberger,Brody1981a,Haller1984a,Berry1987a,Berry1989a}, 
are also discussed.
Finally, we summarize our results in Sec.~\ref{sec-conclusion}.

\section{Models for two coupled KPOs}
\label{sec-model}

The quantum and classical models for the KPO network have been introduced in Ref.~\citenum{Goto2016a}
and well summarized in Ref.~\citenum{Goto2019a}.
In the following, we provide these models in the case of two KPOs.

\subsection{Quantum model}

The quantum model for two KPOs is given by the following Hamiltonian:
\begin{align}
H &= H_1 + H_2 + H_{\mathrm{I}},
\label{eq-H}
\\
H_i &= \hbar \frac{K}{2} a_i^{\dagger 2} a_i^2 - \hbar \frac{p_i}{2} \! \left( a_i^2 + a_i^{\dagger 2} \right) \! 
+ \hbar \Delta a_i^{\dagger} a_i,
\label{eq-Hi}
\\
H_{\mathrm{I}} &= -\hbar \xi_0 \! \left( a_1^{\dagger} a_2 + a_2^{\dagger} a_1 \right),
\label{eq-HI}
\end{align}
where $a_i$ and $p_i$ are the annihilation operator and the parametric pump amplitude, respectively, 
for the $i$th KPO, 
$K$ is the Kerr coefficient, 
$\Delta$ is the detuning of the KPO resonance frequency from half the pump frequency, 
$\xi_0$ is the coupling strength between the two KPOs, 
and $\hbar$ is the reduced Planck constant.

\subsection{Classical model}

The corresponding classical model is derived by replacing the annihilation operator $a_i$ 
with a complex amplitude ${\alpha_i = x_i + \mathrm{i} y_i}$ 
(classical approximation) in the Heisenberg equations of motion for $a_i$~\cite{Goto2016a,Goto2019a}. 
Thus we obtain the equations of motion in the classical model:
\begin{align}
&
\frac{\mathrm{d} x_i}{\mathrm{d} t} = \frac{\partial H_{\mathrm{c}}}{\partial y_i}
= \! \left[ K \! \left( x_i^2 + y_i^2 \right) \! + p_i + \Delta \right] \! y_i - \xi_0 y_j,
\label{eq-x}
\\
&
\frac{\mathrm{d} y_i}{\mathrm{d} t} = -\frac{\partial H_{\mathrm{c}}}{\partial x_i}
= - \! \left[ K \! \left( x_i^2 + y_i^2 \right) \! - p_i + \Delta \right] \! x_i + \xi_0 x_j,
\label{eq-y}
\\
&
H_{\mathrm{c}}(\bm{\mathrm{x}},\bm{\mathrm{y}})
=
\! \sum_{i=1,2} \! \left[ \frac{K}{4} \! \left( x_i^2 + y_i^2 \right)^2 \! 
- \frac{p_i}{2} \! \left( x_i^2 - y_i^2 \right) \! \right.
\nonumber
\\
&
\qquad \qquad \qquad \quad 
\left. + \frac{\Delta}{2} \! \left( x_i^2 + y_i^2 \right) \! \right] \! 
-\xi_0 \! \left( x_1 x_2 + y_1 y_2 \right),
\label{eq-Hc}
\end{align}
where ${j \neq i}$.

For convenience, we introduce the potential energy $V_{\mathrm{c}}(\bm{\mathrm{x}})$ defined by the minimum of 
$H_{\mathrm{c}}(\bm{\mathrm{x}},\bm{\mathrm{y}})$ with respect to $\bm{\mathrm{y}}$, 
which is $H_{\mathrm{c}}(\bm{\mathrm{x}},\bm{0})$ if ${\xi_0 < p_i}$:
\begin{align}
V_{\mathrm{c}}(\bm{\mathrm{x}})
=
\! \sum_{i=1,2} \! \left( \frac{K}{4} x_i^4 - \frac{p_i - \Delta}{2} x_i^2 \right)  \! - \xi_0 x_1 x_2.
\label{eq-V}
\end{align}

\subsection{Parameter setting}

In this work, we take the following values for the above parameters:
\begin{align}
\hbar &= 1,
\label{hbar}
\\
K &= 1,
\\
p_1 &= 3, 
\\
p_2 &= \pi, 
\\
\Delta &= 0,
\label{Delta} 
\\
\xi_0 &= 0,~0.3,~\mathrm{or}~1.
\end{align}
Here ${\hbar = K =1}$ means that the units of energy and frequency are $\hbar K$ and $K$, respectively 
(thus the unit of time is $K^{-1}$).
The pump amplitudes around 3 leads to mean photon numbers around 3, 
because the mean photon numbers for oscillating states are given by ${p_i/K}$ when ${\Delta =0}$~\cite{Goto2016a}.
We choose such small values because we are interested in the dynamics at a few-photon level.
Also, we set $p_2$ to $\pi$ such that $p_2 \simeq p_1$ but $p_2/p_1$ becomes an irrational number, 
because then the ratio between the periods of the two KPOs in the decoupled case (${\xi_0=0}$) is irrational, 
and the dynamics becomes relatively complex.
The three values of $\xi_0$ correspond to regular (integrable), intermediate, and chaotic (nonintegrable) cases, respectively, 
as shown in the next section.
(Note that when ${\xi_0=0}$, that is, the two KPOs are decoupled, 
then the Hamiltonian for each KPO is conserved, and the system is integrable by definition~\cite{comment-integrable}.)

\begin{figure}[ht]
	\includegraphics[width=7cm]{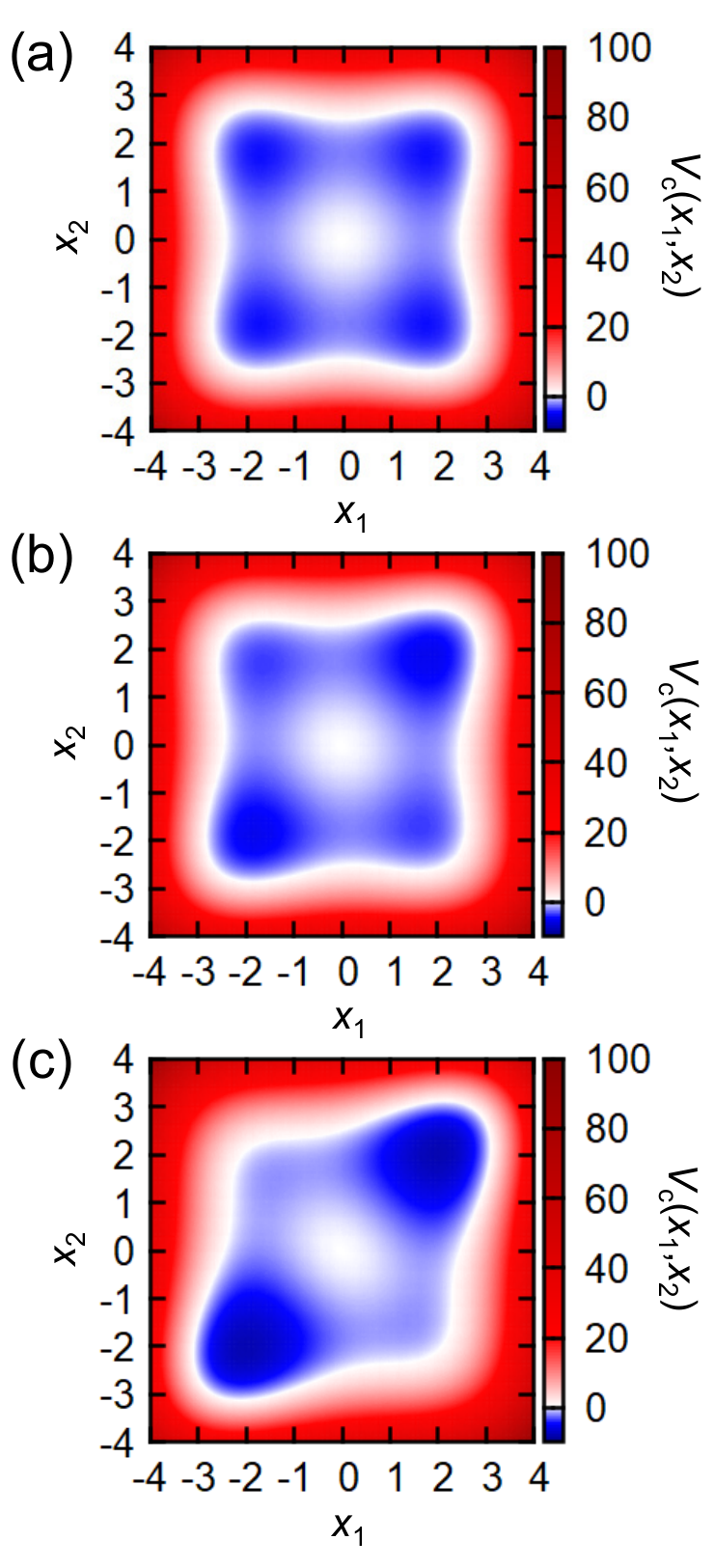}
	\caption{Potential energy $V_{\mathrm{c}}(\bm{\mathrm{x}})$ in Eq~(\ref{eq-V}).
	(a) ${\xi_0=0}$. (b) ${\xi_0=0.3}$. (c) ${\xi_0=1}$. 
	The other parameters are set as Eqs.~(\ref{hbar})--(\ref{Delta}).}
	\label{fig-classical-potential}
\end{figure}

The potential energy $V_{\mathrm{c}}(\bm{\mathrm{x}})$ 
with the above parameters is shown in Fig.~\ref{fig-classical-potential}.
There is a minimum in each quadrant and a maximum at the origin. 
Thus, the origin is unstable. 
In this work, we investigate the dynamics started around the origin
(around the vacuum state in the quantum case).

\section{Chaos in the classical model}
\label{sec-classical}

\subsection{Surface of section (SOS)}

We start with nonintegrability of the classical model.
To check nonintegrability,
the Poincar\'e surface of section (SOS)~\cite{Strogatz,Wimberger} is useful, in particular, 
for systems with two degrees of freedom (four-dimensional phase space), like the present system.
The SOS in the phase space is defined by the section of the energy surface 
${H_{\mathrm{c}}(\bm{\mathrm{x}},\bm{\mathrm{y}})=E}$ ($E$ is a constant) by a plane, e.g., ${y_2=0}$. 
In the case of two degrees of freedom (four-dimensional phase space), 
the object consisting of the intersection points between the SOS and a trajectory, 
which we call the SOS plot, is two-dimensional in general. 
If the system is integrable, however, there is another constant of motion~\cite{comment-integrable}, 
and consequently the SOS plot must be one-dimensional.
Thus we can check whether the system is integrable or not by the dimension of the SOS plot.

\begin{figure}[ht]
	\includegraphics[width=8cm]{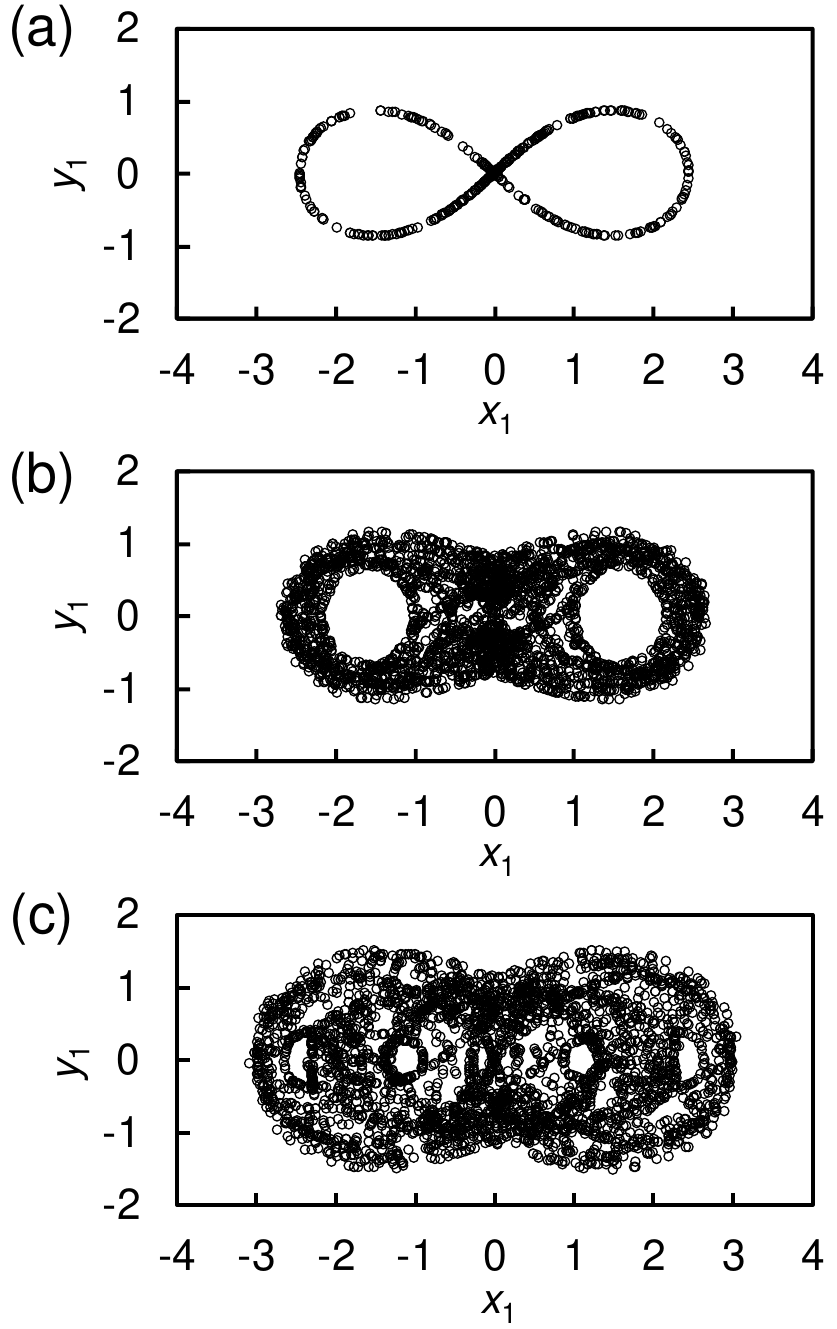}
	\caption{SOS plots in the classical model given by Eqs.~(\ref{eq-x})--(\ref{eq-Hc}). 
	Circles are obtained by plotting ${(x_1,y_1)}$ when trajectories in numerical simulation cross the plane ${y_2=0}$. 
	(a) ${\xi_0=0}$. (b) ${\xi_0=0.3}$. (c) ${\xi_0=1}$. 
	The other parameters are set as Eqs.~(\ref{hbar})--(\ref{Delta}).
	See Appendix~\ref{appendix-classical-SOS} for details.}
	\label{fig-classical-SOS}
\end{figure}

Figure~\ref{fig-classical-SOS} shows the SOS plots in the classical model. 
The SOS plot is one-dimensional~\cite{comment-lemniscate} (integrable and regular) 
in the decoupled case (${\xi_0=0}$) [Fig.~\ref{fig-classical-SOS}(c)], 
two-dimensional (nonintegrable and chaotic)
in the strong-coupling case (${\xi_0=1}$) [Fig.~\ref{fig-classical-SOS}(c)], 
and intermediate in the intermediate case (${\xi_0=0.3}$) [Fig.~\ref{fig-classical-SOS}(b)], as expected.

\subsection{Momentum plot at a minimum of potential (MPMP)}

To demonstrate the nonintegrability more clearly, 
here we introduce another plot, which we call the momentum plot at a minimum of potential (MPMP).
Instead of the plane ${y_2=0}$ for the SOS, 
here we fix the two positions, $x_1$ and $x_2$, at a minimum of the potential $V_{\mathrm{c}}(\bm{\mathrm{x}})$, 
and plot the momenta in the $y_1 y_2$ plane. 
We focus on a potential minimum, because at such a point, the energetically allowable region of the momenta becomes the largest, 
and that will be desirable for visualization.
If the system is integrable, the MPMP must be zero-dimensional (i.e., points), otherwise one-dimensional.

\begin{figure}[hb]
	\includegraphics[width=6.4cm]{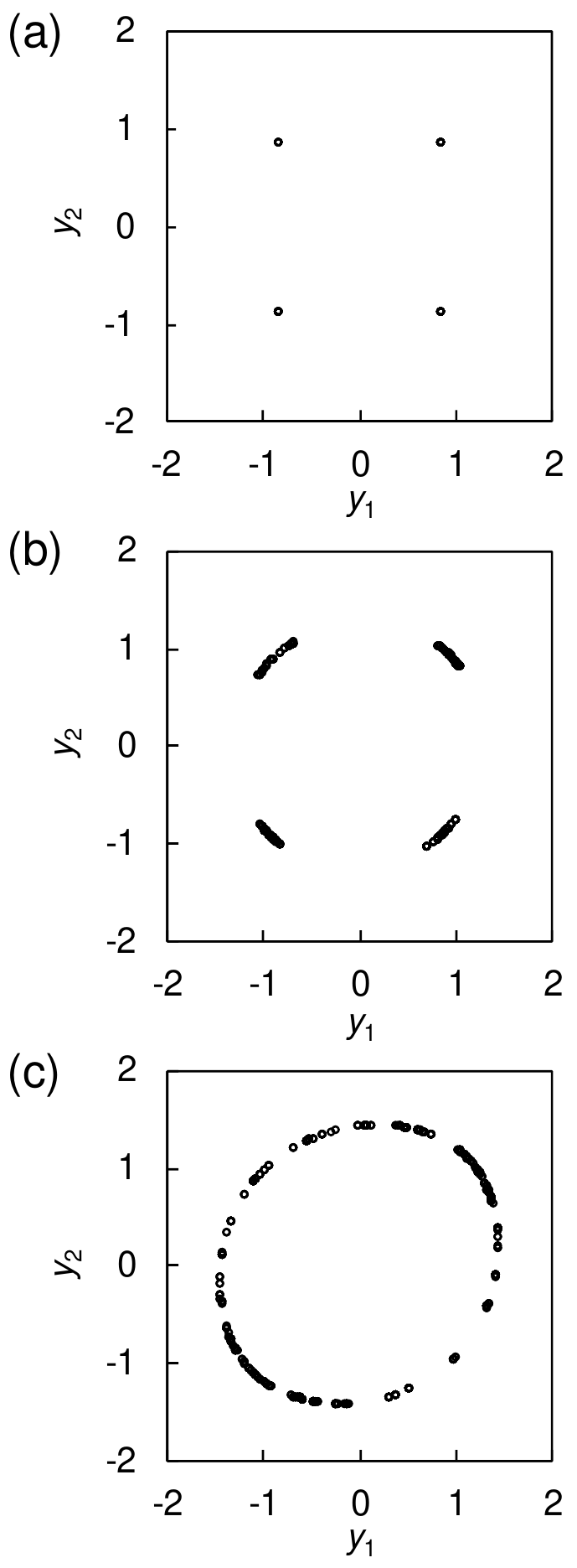}
	\caption{MPMPs in the classical model given by Eqs.~(\ref{eq-x})--(\ref{eq-Hc}). 
	We choose the potential minimum $(X_1,X_2)$ in the first quadrant of the $x_1 x_2$ plane. 
	Circles are obtained by plotting ${(x_1,y_1)}$ when trajectories in numerical simulation pass by close to $(X_1,X_2)$.
	(a) ${\xi_0=0}$ and ${(X_1,X_2)=(1.73,1.77)}$. 
	(b) ${\xi_0=0.3}$ and ${(X_1,X_2)=(1.82,1.85)}$. 
	(c) ${\xi_0=1}$ and ${(X_1,X_2)=(2,2.03)}$. 
	The other parameters are set as Eqs.~(\ref{hbar})--(\ref{Delta}).
	See Appendix~\ref{appendix-classical-MPMP} for details.}
	\label{fig-classical-MPMP}
\end{figure}

Figure~\ref{fig-classical-MPMP} shows the results of the MPMP, where 
we choose the potential minimum in the first quadrant of the $x_1 x_2$ plane.
The MPMP is zero-dimensional for ${\xi_0=0}$ [Fig~\ref{fig-classical-MPMP}(a)], 
one-dimensional for ${\xi_0=1}$ [Fig~\ref{fig-classical-MPMP}(c)], 
and intermediate for ${\xi_0=0.3}$ [Fig~\ref{fig-classical-MPMP}(b)], as expected.
As we will see in Sec.~\ref{sec-quantum}, 
the MPMP is particularly useful in the quantum case, in comparison with the SOS plot.

\subsection{Sensitivity to initial conditions}

\begin{figure}[ht]
	\includegraphics[width=6.1cm]{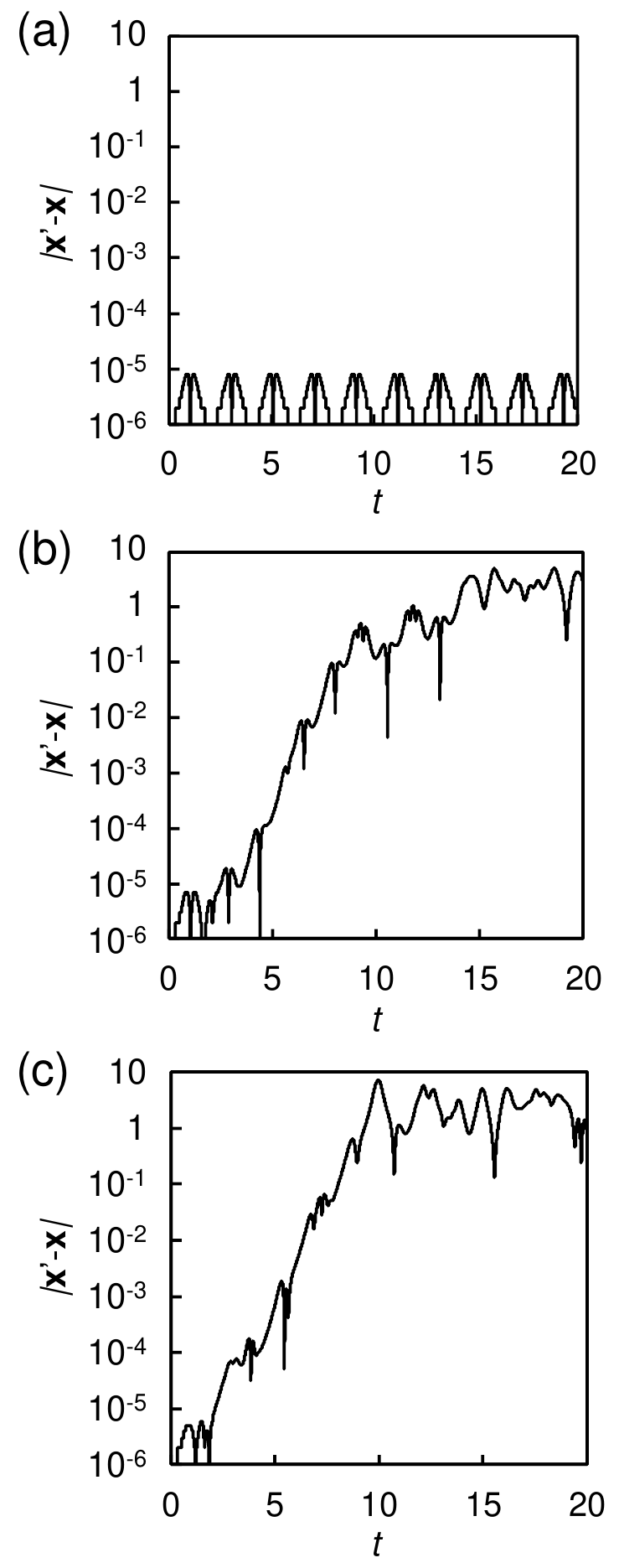}
	\caption{Initial-condition sensitivity in the classical model given by Eqs.~(\ref{eq-x})--(\ref{eq-Hc}). 
	$\bm{\mathrm{x}}(t)$ and $\bm{\mathrm{x}}'(t)$ are two trajectories
	with a very small deviation in their initial conditions. 
	(a) ${\xi_0=0}$. (b) ${\xi_0=0.3}$. (c) ${\xi_0=1}$. 
	The other parameters are set as Eqs.~(\ref{hbar})--(\ref{Delta}).
	See Appendix~\ref{appendix-classical-sensitivity} for details.}
	\label{fig-classical-sensitivity}
\end{figure}

Here we also observe the sensitivity to initial conditions in the classical model.
The results are shown in Fig.~\ref{fig-classical-sensitivity}. 
The Euclidean distance between two trajectories, $\bm{\mathrm{x}}(t)$ and $\bm{\mathrm{x}}'(t)$, 
with a very small deviation in their initial conditions 
saturates at a small value in the integrable case (${\xi_0=0}$), 
but exponentially grows in the nonintegrable cases (${\xi_0=0.3}$ and 1), 
as expected.
(The saturation in the nonintegrable cases comes from the fact that 
the energetically allowable regions are finite.)
This together with the above results for SOS and MPMP 
indicates chaos in the coupled-KPO system.

\section{Quantum signatures of chaos}
\label{sec-quantum}

\subsection{Quantum SOS plots}

Here we introduce quantum versions of the SOS plot.
Instead of plotting intersection points in the classical case, 
we use the time integral of a quasi-probability distribution, the Wigner function or 
the Husimi function (also known as the Q function)~\cite{Milburn,Goto2016a,Wimberger,Leonhardt}, 
with ${y_2=0}$. 
Thus we define the quantum versions of the SOS plot as
\begin{align}
W_{\mathrm{SOS}}(x_1,y_1)
&=
\int_{0}^{T} \! \mathrm{d}t \! \int_{-\infty}^{\infty} \! \mathrm{d}x_2 W(x_1,x_2,y_1,0,t),
\label{eq-WSOS}
\\
Q_{\mathrm{SOS}}(x_1,y_1)
&=
\int_{0}^{T} \! \mathrm{d}t \! \int_{-\infty}^{\infty} \! \mathrm{d}x_2 Q(x_1,x_2,y_1,0,t),
\label{eq-QSOS}
\end{align}
where $W(x_1,x_2,y_1,0,t)$ and $Q(x_1,x_2,y_1,0,t)$ are the Wigner and Husimi functions with ${y_2=0}$ 
for the state vector, $|\psi (t) \rangle$, at time $t$, and $T$ is the final time in each simulation.

Figure~\ref{fig-quantum-SOS} shows the results of the quantum SOS plots corresponding to the classical ones 
in Fig.~\ref{fig-classical-SOS}.
Although the Husimi-type SOS plots shown in Figs.~\ref{fig-quantum-SOS}(a)--\ref{fig-quantum-SOS}(c) 
indicate the classical SOS plots in Figs.~\ref{fig-classical-SOS}(a)--\ref{fig-classical-SOS}(c) to some extent, 
it is hard to distinguish integrability (one-dimensional) from nonintegrability (two-dimensional) 
because of large quantum fluctuations.
(The large fluctuations come from the small mean photon numbers.)
The situation is worse in the Wigner case, as shown in Figs.~\ref{fig-quantum-SOS}(d)--\ref{fig-quantum-SOS}(f), 
because of quantum interference.
(Similar results have been reported for a single driven pendulum~\cite{Lee1993a}.).
This is the reason why we have introduced the MPMP in this work.

\begin{widetext}

\begin{figure}[hbt]
\centering
	\includegraphics[width=17cm]{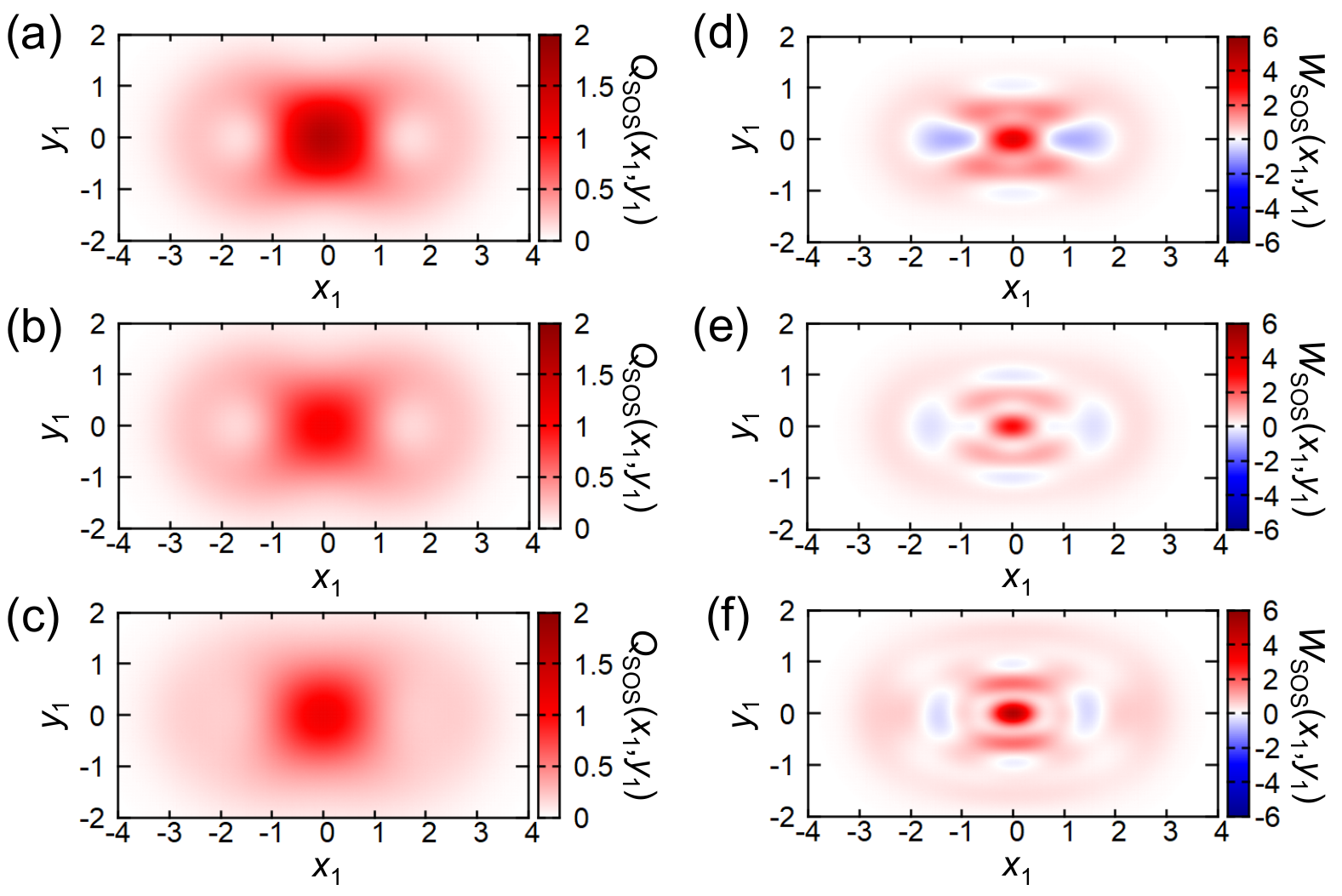}
	\caption{Quantum SOS plots defined by Eqs.~(\ref{eq-WSOS}) and (\ref{eq-QSOS}) 
	in the quantum model given by Eqs.~(\ref{eq-H})--(\ref{eq-HI}). 
	Corresponding classical results are shown in Fig.~\ref{fig-classical-SOS}.
	(a,d) ${\xi_0=0}$. (b,e) ${\xi_0=0.3}$. (c,f) ${\xi_0=1}$. 
	The other parameters are set as Eqs.~(\ref{hbar})--(\ref{Delta}).
	See Appendix~\ref{appendix-quantum-SOS} for details.}
	\label{fig-quantum-SOS}
\end{figure}

\subsection{Quantum MPMPs}

Here we introduce quantum versions of the MPMP as
\begin{align}
W_{\mathrm{MPMP}}(y_1,y_2)
&=
\int_{0}^{T} \! \mathrm{d}t W(X_1,X_2,y_1,y_2,t),
\label{eq-WMPMP}
\\
Q_{\mathrm{MPMP}}(y_1,y_2)
&=
\int_{0}^{T} \! \mathrm{d}t Q(X_1,X_2,y_1,y_2,t),
\label{eq-QMPMP}
\end{align}
where $(X_1,X_2)$ is the position of a minimum of the potential $V_{\mathrm{c}}(\bm{\mathrm{x}})$ in Eq.~(\ref{eq-V}).

Figure~\ref{fig-quantum-MPMP} shows the results of the quantum MPMPs 
together with the corresponding classical results 
in Fig.~\ref{fig-classical-MPMP}.
The Husimi-type MPMPs shown in Figs.~\ref{fig-quantum-MPMP}(a)--\ref{fig-quantum-MPMP}(c) 
clearly indicate the classical MPMPs even with large quantum fluctuations, 
which can be regarded as a quantum signature of chaos (nonintegrability).
The Wigner-type MPMPs shown in Figs.~\ref{fig-quantum-MPMP}(d)--\ref{fig-quantum-MPMP}(f) 
also indicate the classical MPMPs to some extent 
even with quantum interference.
Thus, the MPMP is particularly useful in quantum cases, 
in comparison with the SOS plot.

\begin{figure}[b]
	\includegraphics[width=14cm]{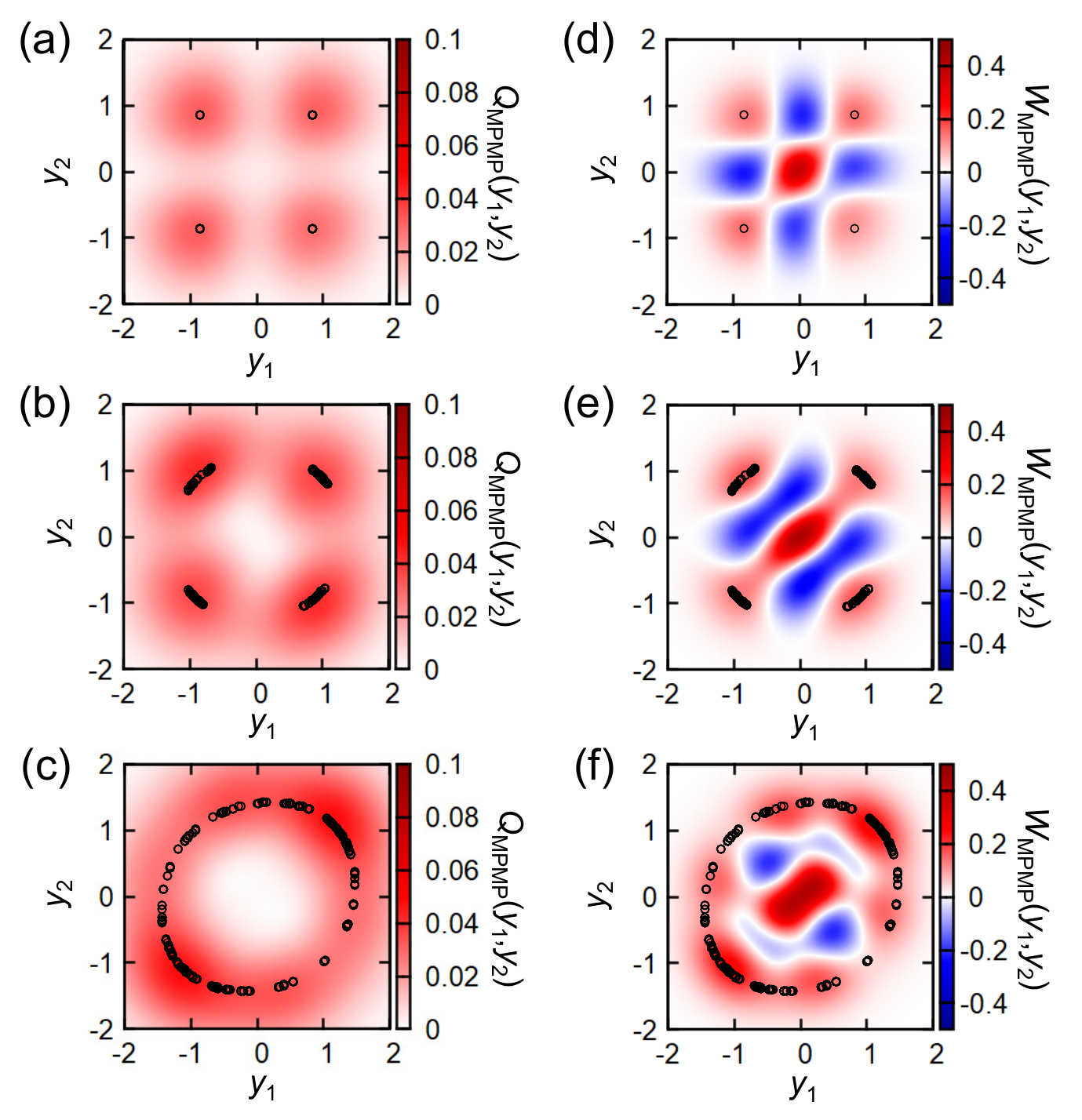}
	\caption{Quantum MPMPs defined by Eqs.~(\ref{eq-WMPMP}) and (\ref{eq-QMPMP}) 
	in the quantum model given by Eqs.~(\ref{eq-H})--(\ref{eq-HI}).
	$(X_1,X_2)$ is set to the position of the potential minimum 
	in the first quadrant of the $x_1 x_2$ plane. 
	Circles represent classical results shown in Fig.~\ref{fig-classical-SOS}.
	(a,d) ${\xi_0=0}$ and ${(X_1,X_2)=(1.73,1.77)}$. 
	(b,e) ${\xi_0=0.3}$ and ${(X_1,X_2)=(1.82,1.85)}$. 
	(c,f) ${\xi_0=1}$ and ${(X_1,X_2)=(2,2.03)}$. 
	The other parameters are set as Eqs.~(\ref{hbar})--(\ref{Delta}).
	See Appendix~\ref{appendix-quantum-MPMP} for details.}
	\label{fig-quantum-MPMP}
\end{figure}

\end{widetext}

\subsection{Out-of-time-ordered correlators (OTOCs)}

Here we discuss the sensitivity to initial conditions in the quantum model. 
In quantum cases, the initial-condition sensitivity can be evaluated by 
the out-of-time-ordered correlators (OTOCs)~\cite{Hashimoto2017a,Fortes2019a,Akutagawa2020a,Hashimoto2020a,Bhattacharyya2021a} 
defined by 
\begin{align}
C_{i,j}(t)
=
-4\langle \psi (0)| [x_i(t), y_j(0)]^2 |\psi (0) \rangle,
\label{eq-quantum-OTOC}
\end{align}
where $|\psi (0) \rangle$ is an initial state vector, 
$[O_1,O_2]=O_1 O_2 - O_2 O_1$ is the commutation relation between two operators, 
$x_i(t)$ is the position operator for the $i$th KPO at time $t$ in the Heisenberg representation, 
and $y_j(0)$ is the initial momentum operator for the $j$th KPO.
The factor of 4 comes from the definitions of the quadrature amplitudes and their commutation relations: 
\begin{align}
x_i(0) = \frac{a_i + a_i^{\dagger}}{2},
\label{eq-x-def}
\\
y_i(0) = \frac{a_i - a_i^{\dagger}}{2\mathrm{i}},
\label{eq-y-def}
\\
[x_i(0),y_i(0)]=\frac{\mathrm{i}}{2}.
\end{align}

The physical meaning of the OTOCs can be extracted by naively replacing the commutator 
with the classical Poisson bracket 
${(\mathrm{i}/2) \{ x_i(t), y_j(0) \}} = {(\mathrm{i}/2) \partial x_i(t)/\partial x_j(0)}$~\cite{Hashimoto2017a}. 
That is, the classical counterpart of $C_{i,j}(t)$, which is denoted by $\tilde{C}_{i,j}(t)$, is given by 
\begin{align}
\tilde{C}_{i,j}(t)
=
\left\langle \left( \frac{\partial x_i(t)}{\partial x_j(0)} \right)^2
\right\rangle,
\label{eq-classical-OTOC}
\end{align}
where $\langle \cdot \rangle$ represents the average over trajectories 
with different initial conditions related to the quantum initial state.
This classical interpretation suggests that the OTOCs are related to the initial-condition sensitivity.

\begin{figure}[ht]
	\includegraphics[width=6.9cm]{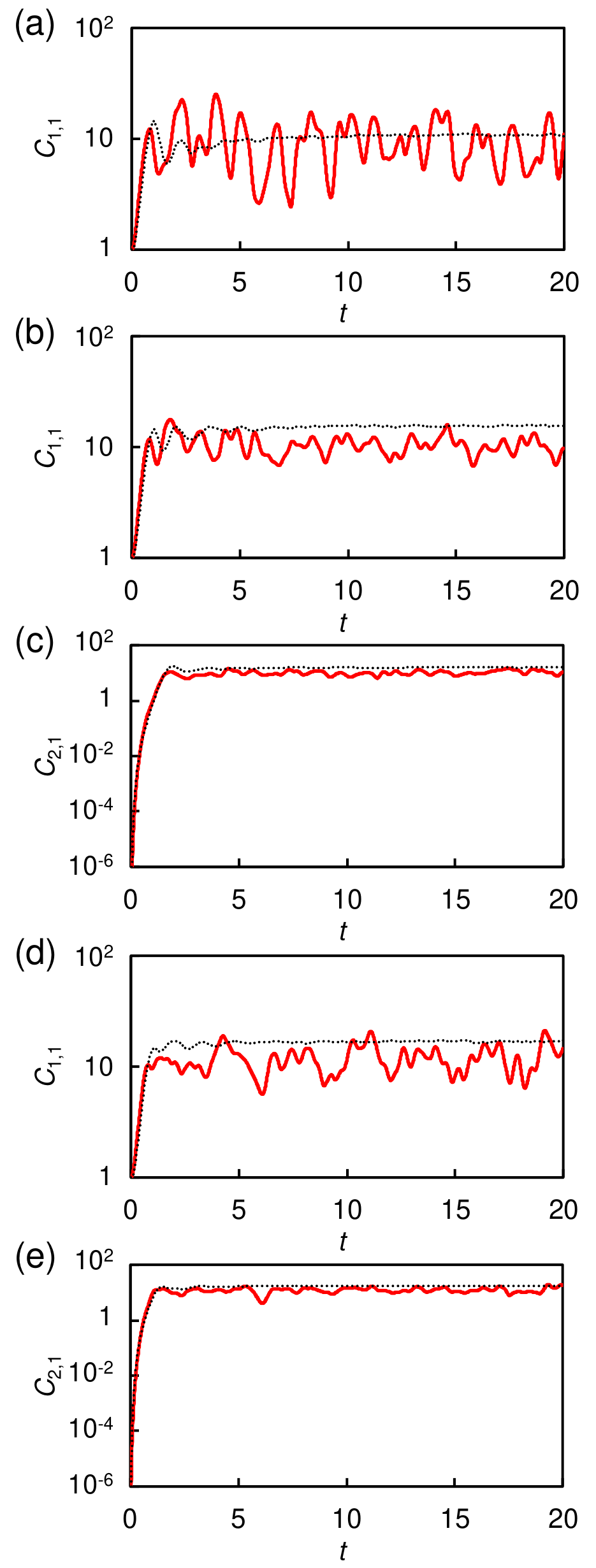}
	\caption{OTOCs in the quantum model (solid lines) 
	and classical counterparts (dotted lines). 
	(a) ${\xi_0=0}$. (b,c) ${\xi_0=0.3}$. (d,e) ${\xi_0=1}$. 
	($C_{2,1}$ for ${\xi_0=0}$ is not shown, because 
	in the decoupled case, $C_{2,1}$ is exactly zero.)
	The other parameters are set as Eqs.~(\ref{hbar})--(\ref{Delta}).
	See Appendix~\ref{appendix-OTOC} for details.}
	\label{fig-OTOC}
\end{figure}

Figure~\ref{fig-OTOC} shows the results of the OTOCs, 
where the solid and dotted lines represent $C_{i,j}(t)$ and $\tilde{C}_{i,j}(t)$, respectively.
First of all, the classical results are in good agreement with the quantum results, 
in particular, around the initial time. 
This indicates that the above classical interpretation of the OTOCs is valid.

However, unlike the classical initial-condition sensitivity shown in Fig.~\ref{fig-classical-sensitivity}, 
the OTOCs rapidly increase only around the initial time and soon saturate. 
This may be due to quantum fluctuations. 
Here it should be noted that the initial rapid increase of the OTOCs does not indicate chaos, 
because this can be seen even in the integrable case, as shown in Fig.~\ref{fig-OTOC}(a). 
Instead, this naturally occurs when the initial state is around an unstable point (maximum of the potential), 
as discussed recently~\cite{Hashimoto2020a,Bhattacharyya2021a}.

On the other hand, 
we can find that the oscillation amplitudes of $C_{1,1}$ in the nonintegrable case (${\xi_0=0.3}$ and 1) shown 
in Figs.~\ref{fig-OTOC}(b) and \ref{fig-OTOC}(d) 
seem smaller than that in the integrable case (${\xi_0=0}$) shown in Fig.~\ref{fig-OTOC}(a).
This difference may be due to more chaotic behavior in the nonintegrable case.
This is another quantum signature of chaos proposed recently~\cite{Fortes2019a}.

\subsection{Energy-level spacing statistics}

Finally, we check a conventional quantum signature of chaos: 
energy-level spacing statistics~\cite{Wimberger,Brody1981a,Haller1984a,Berry1987a,Berry1989a}.
It is known that the energy-level spacing $\Delta_E$ defined by the difference between two neighboring energy levels 
obeys the Poisson distribution (${\propto e^{-\beta \Delta_E}}$) in the integrable (regular) case and 
the Wigner distribution (${\propto \Delta_E e^{-\beta \Delta_E^2}}$) in the nonintegrable  (chaotic) case 
($\beta$ is a constant).
This means that the probability for zero spacing decreases as the system becomes more chaotic, 
which is due to avoided crossings of energy levels induced by complex interactions in chaotic systems~\cite{Wimberger}.

\begin{figure}[ht]
	\includegraphics[width=8cm]{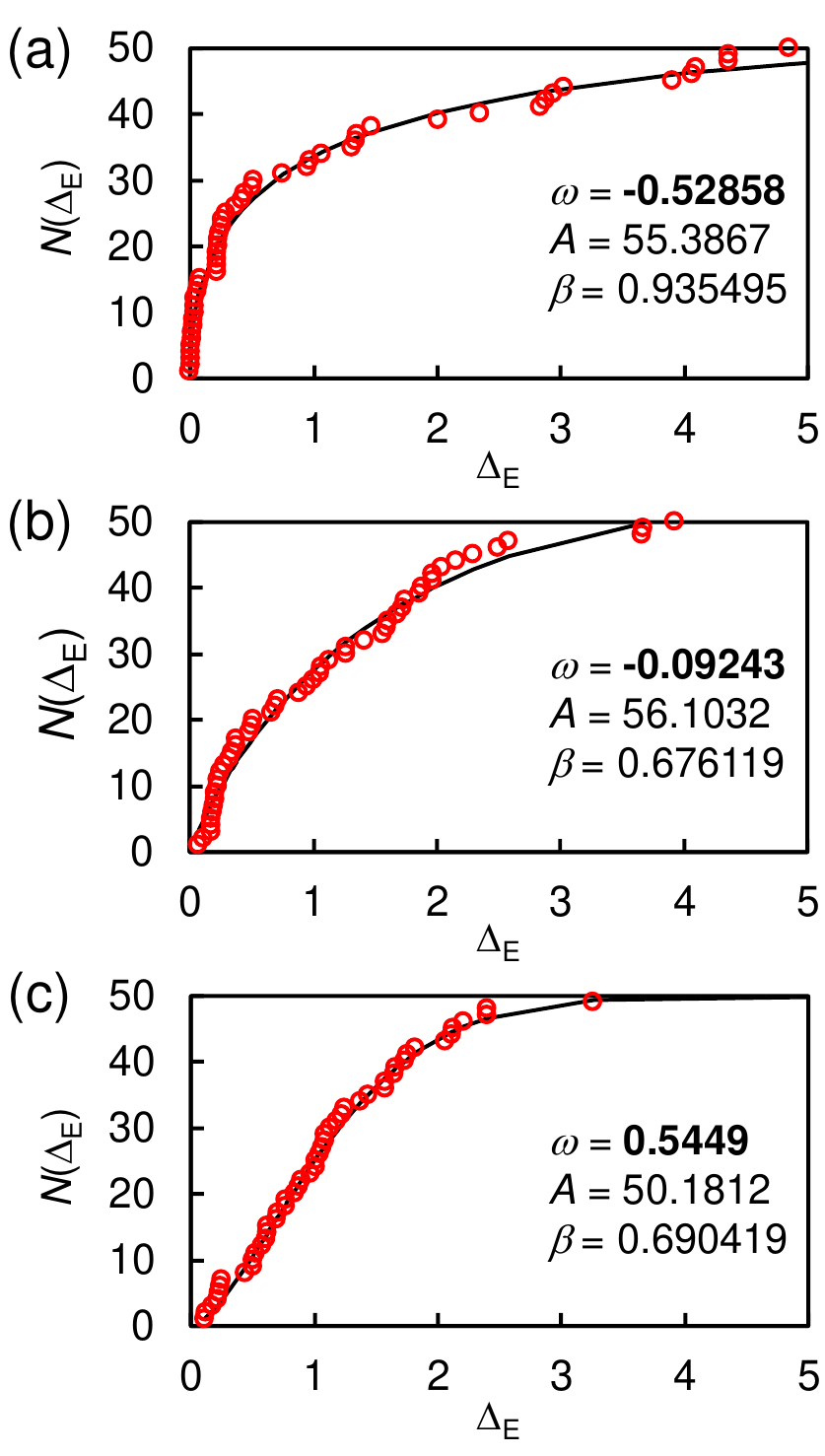}
	\caption{Cumulative energy-level spacing distributions 
	in the quantum model given by Eqs.~(\ref{eq-H})--(\ref{eq-HI}).
	${N(\Delta_E)}$ denotes the number of level spacing equal to or smaller than $\Delta_E$.
	Curves and three parameters ($\omega$, $A$, and $\beta$) 
	are fitting results using the function form ${A (1- e^{-\beta \Delta_E^{\omega+1}})}$. 
	(a) ${\xi_0=0}$. (b) ${\xi_0=0.3}$. (c) ${\xi_0=1}$. 
	The other parameters are set as Eqs.~(\ref{hbar})--(\ref{Delta}).
	See Appendix~\ref{appendix-spacing} for details.}
	\label{fig-spacing}
\end{figure}

The two distributions are unified as 
${\Delta_E^{\omega} e^{-\beta \Delta_E^{\omega+1}}}$~\cite{Brody1981a,Haller1984a}, 
where ${\omega=0}$ and 1 correspond to Poisson and Wigner, respectively.
Note that this distribution can be integrated analytically, 
which leads to the cumulative level spacing distribution ${\propto (1- e^{-\beta \Delta_E^{\omega+1}})}$.
Figure~\ref{fig-spacing} shows the cumulative distributions of the present quantum model 
together with fitting curves using the function form ${A (1- e^{-\beta \Delta_E^{\omega+1}})}$ 
and the fitting results of the three parameters.
The exponent $\omega$ is larger for stronger coupling and 
exceeds 0.5 when ${\xi_0=1}$.
This is another quantum signature of chaos in the coupled-KPO system.

\section{Conclusions}
\label{sec-conclusion}

We have investigated the quantum and classical models for two coupled nondissipative KPOs 
from the viewpoint of chaos.
Using the Poincar\'{e} surface of section (SOS) plot, the momentum plot at a minimum of potential (MPMP) 
(a lower-dimensional version of the SOS plot), 
and the initial-condition sensitivity, 
we have shown that the classical model with nonzero coupling is nonintegrable and hence exhibits chaotic behavior. 
We have also provided quantum signatures of chaos, 
using quantum versions of the SOS plot and the MPMP defined with time integrals of 
the Wigner and Husimi functions, 
out-of-time-ordered correlators (OTOCs), and 
energy-level spacing statistics. 
We have found that the quantum MPMP can distinguish integrability and nonintegrability clearly 
even at a few-photon level (more clearly than the quantum SOS plot), 
and also that the long-term behavior, not the initial behavior, of the OTOCs
can be regarded as a quantum signature of chaos.
The present results suggest that the system of coupled KPOs will be useful 
not only for quantum computing, but also the study on quantum chaos.
We also expect that such an understanding of the nonlinear dynamics in the KPO network 
will be useful for its applications, such as quantum computing and quantum-inspired algorithms~\cite{Goto2019c}.

\section*{Acknowledgments}

This work was supported by JST ERATO (Grant No. JPMJER1601).

\begin{appendix}

\section{SOS plot in the classical model}
\label{appendix-classical-SOS}

To obtain the SOS plots in Fig.~\ref{fig-classical-SOS}, 
we numerically solve Eqs.~(\ref{eq-x})--(\ref{eq-Hc}) 
by the fourth-order Runge-Kutta method with a time step of ${\Delta_t=10^{-4}}$ 
from $t=0$ to 20. 
The initial conditions are set as ${x_1(0)=x_2(0)=0}$ and ${y_i(0)=10^{-6} r_i}$, 
where $r_1$ and $r_2$ are independent random numbers from the standard normal distribution. 
We plot ${(x_1,y_1)}$ when ${y_2 (t) y_2 ({t-\Delta_t}) <0}$, 
which means that the trajectory have crossed the plane ${y_2=0}$.
We iterated the simulation 200 times to obtain enough points.

\section{MPMP in the classical model}
\label{appendix-classical-MPMP}

To obtain the MPMPs in Fig.~\ref{fig-classical-MPMP}, 
we did the same simulation as in the SOS case explained in Appendix~\ref{appendix-classical-SOS}.
We plot ${(y_1,y_2)}$ when ${|\bm{\mathrm{x}} - \bm{\mathrm{X}}|<10^{-3}}$, 
where $\bm{\mathrm{X}}$ denotes the position of the potential minimum 
in the first quadrant of the $x_1 x_2$ plane. 
$\bm{\mathrm{X}}$ is numerically found as 
${\bm{\mathrm{X}}=(1.73,1.77)}$ for ${\xi_0=0}$, 
${\bm{\mathrm{X}}=(1.82,1.85)}$ for ${\xi_0=0.3}$, 
and ${\bm{\mathrm{X}}=(2,2.03)}$ for ${\xi_0=1}$. 
We iterated the simulation $10^5$ times to obtain enough points.

\section{Initial-condition sensitivity in the classical model}
\label{appendix-classical-sensitivity}

To obtain Fig.~\ref{fig-classical-sensitivity},  
we did the same simulation as above with the following initial conditions (no iteration): 
\begin{align}
&
x_1(0)=x_2(0)=x'_2(0)=0, 
\label{eq-sensitivity-initial-x}
\\
&
x'_1(0)=10^{-6}, 
\\
&
y_1(0)=y'_1(0)=0.5\cos (0.65\pi), 
\\
&
y_2(0)=y'_2(0)=0.5\sin (0.65\pi).
\label{eq-sensitivity-initial-y}
\end{align}

\section{Quantum SOS plots}
\label{appendix-quantum-SOS}

Here we give the formulations of $W_{\mathrm{SOS}}(x_1,y_1)$ and $Q_{\mathrm{SOS}}(x_1,y_1)$ 
defined by Eqs.~(\ref{eq-WSOS}) and (\ref{eq-QSOS}) 
in the photon-number basis.

First of all, the Wigner and Husimi functions are defined as~\cite{Goto2016a,Leonhardt} 
\begin{align}
W(\bm{\mathrm{x}},\bm{\mathrm{y}},t)
&=
\left( \frac{2}{\pi} \right)^2 \! 
\mathrm{tr} \! \left[ D^{(1)} (2\alpha_1 ) P_1 D^{(2)} (2\alpha_2 ) P_2 {\rho (t)} \right],
\\
Q(\bm{\mathrm{x}},\bm{\mathrm{y}},t)
&=
\frac{1}{\pi^2} {\langle \alpha_1 |} {\langle \alpha_2 |} {\rho (t)} {| \alpha_1 \rangle} {| \alpha_2 \rangle},
\end{align}
where ${\alpha_i = x_i + \mathrm{i} y_i}$, 
${D^{(i)} (\alpha ) =e^{\alpha a_i^{\dagger} - \alpha^* a_i}}$ is the so-called displacement operator, 
${P_i = e^{\mathrm{i} a_i^{\dagger} a_i}}$ is the parity operator, 
${|\alpha_i \rangle = D^{(i)} (\alpha_i ) |0\rangle}$ is a coherent state ($|0\rangle$ is the vacuum state), and
${\rho  (t) = |\psi (t) \rangle \langle \psi (t)|}$ is the density operator corresponding to the state vector 
$|\psi (t) \rangle$.

In the photon-number basis $\{ |n_1 \rangle |n_2 \rangle \}$,
we obtain~\cite{Goto2016a}
\begin{widetext}
\begin{align}
W(\bm{\mathrm{x}},\bm{\mathrm{y}},t)
&=
\left( \frac{2}{\pi} \right)^2
\sum_{m_1=0}^{\infty} \sum_{m_2=0}^{\infty} \sum_{n_1=0}^{\infty} \sum_{n_2=0}^{\infty}
 (-1)^{n_1+n_2} 
D^{(1)}_{m_1,n_1} (2\alpha_1 ) D^{(2)}_{m_2,n_2} (2\alpha_2 ) \psi_{n_1,n_2}(t) \psi^*_{m_1,m_2}(t),
\label{eq-W-number}
\\
Q(\bm{\mathrm{x}},\bm{\mathrm{y}},t)
&=
\frac{1}{\pi^2} 
\sum_{m_1=0}^{\infty} \sum_{m_2=0}^{\infty} \sum_{n_1=0}^{\infty} \sum_{n_2=0}^{\infty}
{\langle \alpha_1 | n_1 \rangle} {\langle \alpha_2 | n_2 \rangle} \psi_{n_1,n_2}(t)  \psi^*_{m_1,m_2}(t)
{\langle m_1 | \alpha_1 \rangle} {\langle m_2 | \alpha_2 \rangle},
\label{eq-Q-number}
\end{align}
with
\begin{align}
&
\psi_{n_1,n_2}(t) = \langle n_1| \langle n_2| \psi (t) \rangle,
\label{eq-psi-number}
\\
&
D^{(i)}_{m_i,n_i} (\alpha)
=
e^{-|\alpha |^2/2} \sqrt{m_i ! n_i !} \sum_{k=0}^{\mathrm{min}(m_i,n_i)} 
\! \frac{1}{k!} \frac{\alpha^{m_i - k}}{(m_i-k)!} \frac{(-\alpha^*)^{n_i-k}}{(n_i-k)!},
\label{eq-D-number}
\\
&
{\langle n_i | \alpha_i \rangle} = \frac{\alpha_i^{n_i}}{\sqrt{n_i !}} e^{-|\alpha_i |^2/2}.
\label{eq-alpha-number}
\end{align}

The integrals $\int_{-\infty}^{\infty} \! \mathrm{d}x_2 W(x_1,x_2,y_1,0,t)$ and 
$\int_{-\infty}^{\infty} \! \mathrm{d}x_2 Q(x_1,x_2,y_1,0,t)$ 
required for $W_{\mathrm{SOS}}(x_1,y_1)$ and $Q_{\mathrm{SOS}}(x_1,y_1)$ 
are obtained by Eqs.~(\ref{eq-W-number})--(\ref{eq-alpha-number}) together with the following formulae:
\begin{align}
&
\int_{-\infty}^{\infty} \! \mathrm{d}x_2 D^{(2)}_{m_2,n_2} (2x_2)
=
\sqrt{\frac{\pi}{2} m_2 ! n_2 !} \sum_{k=0}^{\mathrm{min}(m_2,n_2)} 
\! \frac{(-1)^{n_2-k} (m_2 + n_2 -2k -1)!!}{k! (m_2-k)! (n_2-k)!} 
\delta_{\mathrm{even}} (m_2+n_2),
\\
&
\int_{-\infty}^{\infty} \! \mathrm{d}x_2 \langle \alpha_2|m_2 \rangle \langle n_2| \alpha_2 \rangle
=
\sqrt{\frac{\pi}{m_2! n_2!}} \frac{(m_2+n_2-1)!!}{2^{(m_2+n_2)/2}} \delta_{\mathrm{even}} (m_2+n_2),
\end{align}
\end{widetext}
where ${\delta_{\mathrm{even}} (n)=1}$ if $n$ is even, otherwise ${\delta_{\mathrm{even}} (n)=0}$.

The numerical results of 
$W_{\mathrm{SOS}}(x_1,y_1)$ and $Q_{\mathrm{SOS}}(x_1,y_1)$ shown in Fig.~\ref{fig-quantum-SOS} 
are obtained by accumulating the integrals $\int_{-\infty}^{\infty} \! \mathrm{d}x_2 W(x_1,x_2,y_1,0,t)$ and 
$\int_{-\infty}^{\infty} \! \mathrm{d}x_2 Q(x_1,x_2,y_1,0,t)$ multiplied by a time step of ${\Delta_t=10^{-3}}$
from ${t=0}$ to ${T=20}$. 
Here $\psi_{n_1,n_2} (t)$ necessary for the integrals is obtained by solving the Schr\"{o}dinger equation 
with the Hamiltonian in Eq.~(\ref{eq-H}) in the photon-number basis 
by the fourth-order Runge-Kutta method with the time step of ${\Delta_t=10^{-3}}$, 
the initial state set to the vacuum state, and 
the maximum photon number of 30.

\section{Quantum MPMPs}
\label{appendix-quantum-MPMP}

The numerical results of 
$W_{\mathrm{MPMP}}(y_1,y_2)$ and $Q_{\mathrm{MPMP}}(y_1,y_2)$ shown in Fig.~\ref{fig-quantum-MPMP} 
are obtained by accumulating the Wigner and Husimi functions multiplied by a time step of ${\Delta_t=10^{-3}}$ 
from ${t=0}$ to ${T=20}$, 
where the Wigner and Husimi functions are obtained by using Eqs.~(\ref{eq-W-number})--(\ref{eq-alpha-number}) 
with $\psi_{n_1,n_2} (t)$ obtained by the same simulation as in Appendix~\ref{appendix-quantum-SOS}.
$(X_1,X_2)$ is numerically found,
as mentioned in Appendix~\ref{appendix-classical-MPMP}.

\section{OTOC}
\label{appendix-OTOC}

The results shown in Fig.~\ref{fig-OTOC} are obtained as follows.

Using ${x(t)=e^{\mathrm{i} Ht} x(0) e^{-\mathrm{i}Ht}}$,
the OTOC defined by Eq.~(\ref{eq-quantum-OTOC}) is 
formulated in the eigenenergy basis $\{ |E_k \rangle \}$ as~\cite{Hashimoto2017a}
\begin{widetext}
\begin{align}
&
C_{i,j}(t)=
\sum_{k=0}^{\infty} \sum_{m=0}^{\infty} \sum_{n=0}^{\infty} 
\langle \psi (0)|E_k \rangle \langle E_k|[x_i(t),y_j(0)]|E_m \rangle
\langle E_m |[x_i(t),y_j(0)]|E_n \rangle
\langle E_n|\psi (0) \rangle,
\\
&
\langle E_k|[x_i(t),y_j(0)]|E_m \rangle
=
\sum_{l=0}^{\infty} 
\left[
e^{\mathrm{i} (E_k - E_l)t} \langle E_k| x_i(0) |E_l \rangle
\langle E_l| y_j(0) |E_m \rangle
-
e^{\mathrm{i} (E_l - E_m)t} \langle E_k| y_j(0) |E_l \rangle
\langle E_l| x_i(0) |E_m \rangle
\right].
\end{align}
\end{widetext}
Thus, we can obtain $C_{i,j}(t)$ by using $\{ |E_k \rangle \}$ 
obtained by numerically diagonalizing the Hamiltonian in the photon-number basis
with the maximum photon number of 30, 
the same as the above simulation, 
and using the definitions of $x_i(0)$ and $y_i(0)$ in Eqs.~(\ref{eq-x-def}) and (\ref{eq-y-def}).
For the comparison with the results in Fig.~\ref{fig-classical-sensitivity}, 
the initial state is set to coherent states 
as ${|\psi (0) \rangle = |\alpha^{(0)}_1 \rangle |\alpha^{(0)}_2 \rangle}$, 
where ${\alpha^{(0)}_1=x^{(0)}_1 + \mathrm{i} y^{(0)}_1}$ 
[${x^{(0)}_1=0}$ and ${y^{(0)}_1=0.5 \cos (0.65\pi)}$] and 
${\alpha^{(0)}_2=x^{(0)}_2 + \mathrm{i} y^{(0)}_2}$ 
[${x^{(0)}_2=0}$ and ${y^{(0)}_2=0.5 \sin (0.65\pi)}$], 
which correspond to the initial conditions for Fig.~\ref{fig-classical-sensitivity} 
in Eqs.~(\ref{eq-sensitivity-initial-x})--(\ref{eq-sensitivity-initial-y}).
The resultant  $C_{i,j}(t)$ are shown by the solid lines in Fig.~\ref{fig-OTOC}.

For the classical counterparts, 
we calculate two trajectories, $\bm{\mathrm{x}}(t)$ and $\bm{\mathrm{x}}'(t)$, 
with the following initial conditions:
\begin{align}
&
x_1(0)=x^{(0)}_1 + \Delta_x r_1, 
\label{eq-OTOC-initial-x}
\\
&
x'_1(0)=x_1(0)+\delta_x, 
\\
&
x_2(0)=x'_2(0)=x^{(0)}_2 + \Delta_x r_2, 
\\
&
y_1(0)=y'_1(0)=y^{(0)}_1 + \Delta_y r_3, 
\\
&
y_2(0)=y'_2(0)=y^{(0)}_2 + \Delta_y r_4,
\label{eq-OTOC-initial-y}
\end{align}
where $r_j$ (${j=1, \ldots, 4}$) are independent random numbers from 
the standard normal distribution, 
${\Delta_x = \Delta_y =0.5}$ correspond to the standard deviations 
of the quantum fluctuations of $x_i$ and $y_i$, and
${\delta_x = 0.5}$ is the deviation for the evaluation of the partial derivative 
in Eq.~(\ref{eq-classical-OTOC}).
(The large $\delta_x$ comparable to the quantum fluctuations 
is used to mimic the saturation of the OTOCs in the quantum model.)
Using the two trajectories, $\tilde{C}_{i,1}(t)$ shown by the dotted lines in Fig.~\ref{fig-OTOC}
are obtained as follows:
\begin{align}
\tilde{C}_{i,1}(t)
=
\left\langle \left( \frac{x'_i(t) - x_i(t)}{\delta_x} \right)^2
\right\rangle,
\end{align}
where the average was taken over $10^4$ iterations.

\section{Energy-level spacing statistics}
\label{appendix-spacing}

The results in Fig.~\ref{fig-spacing} are obtained as follows.
First, we numerically diagonalize the Hamiltonian in Eq.~(\ref{eq-H}) 
in the photon-number basis with the maximum photon number of 30, 
the same as above.
Then, taking the parity invariance of the Hamiltonian into account, 
we classify the energy eigenstates into two groups with even and odd total photon numbers.
Here we focus on the even eigenstates and sort the corresponding energies in ascending order. 
Thus we obtain the energy-level spacing as the difference between two neighboring energies.
To avoid the effects of the finite photon numbers, 
we take 50 spacings from the smallest, which are plotted in Fig.~\ref{fig-spacing}.
The curves in Fig.~\ref{fig-spacing} are obtained by fitting ${A (1- e^{-\beta \Delta_E^{\omega+1}})}$ 
to the 50 points, where $\omega$, $A$, and $\beta$ are fitting parameters.

\end{appendix}


\begin{thebibliography}{99}
\bibitem{comment-integrable}
A Hamiltonian system with $n$ degrees of freedom ($2n$-dimensional phase space) 
is integrable if there are $n$ constants of motion~\cite{Wimberger}.
\bibitem{comment-lemniscate}
This curve is known as the lemniscate of Bernoulli.
\end{thebibliography}
\end{document}